\pdfoutput=1

\documentclass[reprint,aps,prd,superscriptaddress,showpacs,preprintnumbers,amsmath,amssymb,floatfix]{revtex4-2} 

\usepackage[euler]{textgreek}
\usepackage[mathlines]{lineno}
\usepackage{graphicx,subfigure}
\usepackage{epstopdf}
\usepackage{dcolumn}
\usepackage{bm}
\usepackage{color}
\usepackage{upgreek}
\usepackage{afterpage}
\usepackage{tikz, pgfplots}
\usepackage[usestackEOL]{stackengine}
\pgfplotsset{compat=1.17}
\usepackage[version=4]{mhchem} 
\usepackage[unicode,colorlinks=true]{hyperref}
\usepackage{soul}
\usepackage{makecell}

\newcommand{\cevns}{\ensuremath{\text{CE\textnu{}NS}}}
\newcommand{\fluxunits}{{n/cm$^2$s}}

\definecolor{kittycolor}{RGB}{168,168,244}

\begin{document}

\preprint{ \today}

\title{Neutron capture-induced nuclear recoils as background for \cevns~measurements at reactors}
\author{A.J.~Biffl}  \email{Corresponding author: alexander.biffl@ucdenver.edu}\affiliation{Department of Physics, University of Colorado Denver, Denver, Colorado 80217, USA}
\author{A.~Gevorgian}  \affiliation{Department of Physics, University of Colorado Denver, Denver, Colorado 80217, USA}
\author{K.~Harris}  \affiliation{Department of Physics, University of Colorado Denver, Denver, Colorado 80217, USA}
\author{A.N.~Villano} \email{Corresponding author: anthony.villano@ucdenver.edu}  \affiliation{Department of Physics, University of Colorado Denver, Denver, Colorado 80217, USA}

\smallskip
\date{\today}

\noaffiliation


\smallskip 

\sethlcolor{green}
\newcommand{\hlh}[1]{#1} 

\begin{abstract}
Nuclear reactors represent a promising neutrino source for \cevns\ (coherent-elastic
neutrino-nucleus scattering) searches. However, reactor sites also come with high ambient neutron
flux. Neutron capture-induced nuclear recoils can create a spectrum that strongly overlaps the
\cevns\ signal for recoils $\lesssim$\,100\,eV for nuclear reactor measurements in silicon or
germanium detectors.  This background can be particularly critical for low-power research reactors
providing a moderate neutrino flux. In this work we quantify the impact of this background and
show that, for a measurement 10\,m from a 1\,MW reactor, the effective thermal neutron flux should
be kept below $\sim$~7$\times$~10$^{-4}$\,\fluxunits\ so that the \cevns\ events can be measured
at least at a 5$\sigma$ level with germanium detectors in 100~kg\,yr exposure time.  This flux
corresponds to 60\% of the sea-level flux but needs to be achieved in a nominally high-flux
(reactor) environment.  Improved detector resolution can help the measurements, but the thermal
flux is the key parameter for the sensitivity of the experiment.  For silicon detectors, the
constraint is even stronger and thermal neutron fluxes must be near an order of magnitude lower.
This constraint highlights the need of an effective thermal neutron mitigation strategy for future
low threshold \cevns\ searches. \hlh{In particular, the neutron capture-induced background can be efficiently
reduced by active veto systems tagging the deexcitation gamma
following the capture.}

\end{abstract}


\maketitle

%
%
%
%
%

%
%
%
%
%
%

\section{\label{sec:intro}Introduction}

Coherent Elastic Neutrino Nucleus Scattering (\cevns) is the transfer of momentum between a neutrino
and a nucleus as a whole via neutral current exchange. This process is experimentally significant for detecting low energy ($\lesssim$~MeV) neutrinos due to its large cross section, 
two to four orders of magnitude larger than the commonly used inverse beta decay (IBD)~\cite{Oralbaev_2016, Baxter_2020}.
This allows \cevns\ detectors to be much smaller than their IBD counterparts.

The \cevns\ process has already been measured at an accelerator site by the COHERENT Collaboration using the Spallation
Neutron Source at Oak Ridge National Laboratory as a neutrino source~\cite{doi:10.1126/science.aao0990}. The accelerator provides
neutrinos of a few tens of MeV, and other measurements using this type of source are planned, for
example the upcoming European Spallation Source~\cite{Baxter_2020}.  Likewise, recent studies show
potential for measurements at reactor sites, where the sources of MeV antineutrinos are commercial or
research fission
reactors~\cite{colaresi2021reactorcevnsobservation,PhysRevLett.129.211802,CONNIE:2021ggh,PhysRevLett.126.041804,PhysRevD.106.L051101,NUCLEUS:2019igx,https://doi.org/10.48550/arxiv.2111.06745,MINER2021SiDetectors}.

The trade-off for \cevns' relatively high interaction rate is that it is exceptionally difficult
to detect because of the tiny (sub-keV) recoil of the target nuclei, even despite recent advances
in detector resolution near threshold, for example in
silicon~\cite{doi:10.1063/1.5010699,PhysRevLett.119.131802}. One major hurdle is that
distinguishing a nuclear recoil event caused by \cevns\ from one caused by other sources is
difficult. In the vicinity of nuclear reactors, a large reactogenic thermal neutron flux could cause enough
neutron-capture-induced recoil events to hide the \cevns\ signal. A capture of a neutron via the (n,$\gamma$) process produces one or more
$\sim$MeV scale prompt gamma rays and a prompt nuclear recoil at the sub-keV scale --- coinciding
with the energy scale of events from reactor neutrinos.  

The recoil spectrum from neutron capture on silicon has been recently
measured~\cite{PhysRevD.105.083014} and creates nuclear recoil events down to $\sim$100\,eV.  This
paper investigates the thermal neutron background mitigations needed to measure a \cevns\ signal
in the vicinity of a nuclear reactor.

\section{\label{sec:cevns_spec} \texorpdfstring{CE\MakeLowercase{\textnu{}}NS and capture spectra}{CEnNS and capture spectra}}

We used the model given by Mueller~\cite{PhysRevC.83.054615} for the four principle fissionable
isotopes as the model for reactor antineutrino emissions.  We attempted to emulate the expected
emissions of a low-enriched ($\sim20\%$ $^{235}$U) reactor fuel composition, i.e. that employed by
the MINER Collaboration at the Nuclear Science Center at Texas A\&M
University~\cite{AGNOLET201753}. We assume neutrino emissions consistent with the MINER
study~\cite{MINER2021SiDetectors}, i.e., dominated by $^{235}$U (96.7\%) with smaller fractions of
$^{238}$U and $^{239}$Pu (1.3 and 2.0\% respectively), and negligible ($\lesssim0.1$\%)
$^{241}$Pu.

We tested the target materials silicon and germanium with natural isotope distributions.  The
reactor spectrum was convoluted with the known \cevns\ cross section~\cite{Baxter_2020} using the
form factor model given by~\cite{PhysRevC.60.014903} to yield a recoil energy spectrum for \cevns\
events. The form factor was quite close to 1, with a largest deviation from Ge at a recoil energy
of around 1\,keV, where it had a value of 0.9895.

To calculate the expected recoil energy spectrum from neutron captures, we use
\texttt{nrCascadeSim}, a publicly available dedicated simulation tool developed by two of the
authors here for nuclear recoils resulting from neutron captures~\cite{Villano2022}. This software
was used to generate a large sample of capture-induced recoil energies from which we construct a
probability density function (PDF) to sample. The code takes into account the most probable
multistep cascades in the deexcitation process following capture and the possibility of
decay-in-flight for the intermediate atoms/ions within the cascade.

\section{\label{sec:thingiforgot}Assumed detector configuration}

Recoil datasets were simulated with a total of 3,000 (10,000) \cevns\ events, corresponding to
approximately a 100~kg\,yr exposure time of natural silicon (germanium) detectors, at the \cevns\
event rate calculated by~\cite{MINER2021SiDetectors} for detector deployment 10~m from the reactor
core. Events that produce energies below the detection threshold are removed when applying the
resolution model.  Detectors were modeled with a range of different effective resolutions (defined
here as resolution $\sigma$ at a recoil energy of 50\,eV).  All detector resolutions obeyed a
resolution function of the form:

\begin{equation}\label{eq:res}
    \sigma(E) = \sqrt{\sigma_0^2 + A E}
\end{equation}

\noindent where $E$ is the recoil energy, $\sigma_0$ is the baseline resolution, and $A$ is a
detector-specific factor which is varied to achieve a given effective resolution. We also consider
several values of $\sigma_0$: 1\,eV, 5\,eV, 10\,eV, and 25\,eV. Table~\ref{tab:current_detectors}
shows resolution criteria that are either future targets or have already been achieved for several
experimental efforts. We can see that the current best demonstrated solid-target baseline
resolution comes from the $\nu$-cleus Collaboration at
3.7\,eV~\cite{strauss2017vCleusDetectors}--between our two lowest baseline resolution points and
probably achievable in the near future for Si and Ge. Going by the table it is seen that a 25\,eV
baseline resolution is probably on the horizon for many detectors. For the solid detectors in the
table we also see that the typical threshold is a factor of 3--9 larger than the baseline
resolution--roughly in line with our requirement that events have energies $\geq$5$\sigma_0$ (see
below). 

\begin{table}[!hbt]
    \bigskip 
    \centering
    \vskip-\abovecaptionskip
    \vskip+\belowcaptionskip
    \begin{tabular}{l c c c}
        \hline
        \hline
        Experiment & \makecell{Detector\\ type} & Threshold & \makecell{Baseline\\ resolution}\\
        \hline
        CONNIE\cite{aguilar2019CONNIE_results,moroni20156CONNIE_detectors} & Si CCD & 28~eVee\footnotemark[2] & 5.5 eVee\footnotemark[2]\\
        
        CONUS\cite{bonet2022nuMagMoment4} & Ge PPC & 200~eVee\footnotemark[2] & 25 eVee\footnotemark[2]\\

        Dresden-II\cite{colaresi2021reactorcevnsobservation} & Ge PPC & 200~eVee\footnotemark[2] & 33 eVee\footnotemark[2]\\
        
        MINER\cite{MINER2021SiDetectors} & Si/Ge cal & $\sim20$~eV\footnotemark[1] & 5 eV\footnotemark[1] \\
        
        RED-100\cite{Akimov2019CENNS_review,akimov2020Red100}  & Xe 2PS & 300~eV\footnotemark[1] & $\cdots$\\
        RICOCHET\cite{formaggio2012Ricochet,Leder2018ricochetsim} & Si/Ge bol & $\sim50$~eV\footnotemark[1] & 17~eV\footnotemark[2] \\
        TEXONO\cite{Singh2017TEXONO_background,singh2019TEXONO_DM_limits} & Ge PPC & 300-400~eVee\footnotemark[1] & ~45 eVee\footnotemark[2]\\

        $\nu$-cleus\cite{strauss2017vCleusDetectors} & \Centerstack{CaWO$_4$ /\\ Al$_2$O$_3$ cal} & 19.7~eV\footnotemark[2] & 3.7~eV\footnotemark[2]\\
        
        
        $\nu$GeN\cite{Belov2015NuGen}  & Ge PPC & $350$~eV\footnotemark[2] & 93 eV\footnotemark[2]\\

        \hline\hline
        
    \end{tabular}
    \caption{Summary of \cevns\ searches at reactors. 
Nonstandard abbreviations used are as follows. \textbf{cal}:~Calorimeter. \textbf{2PS}:~Two-phase scintillator. \textbf{bol}:~Bolometer. Note also eVee refers to electron-equivalent (ionization) energy; other energies are heat.}
    \footnotetext[1]{target}
    \footnotetext[2]{achieved}
    \label{tab:current_detectors}
\end{table}

Different levels of ambient thermal neutron flux were surveyed between about $\sim 10^{-7}$~\fluxunits and
$\sim 10^{-3}$~\fluxunits. The flux range is motivated by where this analysis produces
high significance (5$\sigma$) in the presence of modest ``other'' background (see below). The flux
range includes the approximate behavior of the
Dresden-II~\cite{colaresi2021reactorcevnsobservation} published thermal neutron
flux --- 0.25~\fluxunits --- if we account for an approximately 3000 times higher reactor output power
of that commercial reactor.  The fluxes were translated to a total capture event count via the
neutron capture cross section with natural silicon and germanium and assuming an exposure of
100~kg\,yr. The capture events were then sampled from the output PDF of \texttt{nrCascadeSim v1.4.2}. All
sampled recoils with energy $\leq5\sigma_0$ were eliminated to emulate data collection with a
finite energy threshold event trigger. 

Thermal neutrons are not the only nuclear recoil backgrounds possible in \cevns\ experiments, so
we include an ``other'' nuclear recoil background with shape $\propto E^{-0.9}$, consistent with
neutron-scatter background data given by \cite{AGNOLET201753}, and very similar to the $E^{-1.2}$
form mentioned in \cite{PhysRevLett.129.211802}. From MINER's 2017 background measurement they
expect a combined 100\,events/kg/day in nuclear recoil and electron recoil backgrounds but only
5-20\,events/kg/day in \cevns\ events~\cite{AGNOLET201753}. The MINER authors acknowledge the
need to control background and other experiments have demonstrated backgrounds with between
40-60\% of what is expected from the \cevns\
process~\cite{https://doi.org/10.48550/arxiv.2111.06745,Billard_2017}. Given this information we
normalize our nuclear recoil background to 60\% of the \cevns\ rate below about 1\,keV. We do not
add any further electron recoil background because it is often diluted to higher electron
equivalent energies in ionization-sensitive detectors. Furthermore, the electron recoils in some
cases can be discriminated~\cite{NeogPhonon2022,LeeNaITl2015,Wei_2016}.  \hlh{This background model is expected
to represent approximately the maximum acceptable background to a typical \protect{\cevns}\ search.}
Other variations of our fits have been completed in
the release of our code for this analysis~\cite{Biffl_Harris_Villano_2023}. 

\section{\label{sec:comp}Statistical Comparison}

The combined sample of recoil energies from the three sources (\cevns, neutron capture, and other 
backgrounds), were separated back into \cevns\ and neutron-capture components by binned maximum-likelihood fits with the likelihood function:

\begin{equation}
    \label{eqn:likelihood_alt}
    \mathcal{L} = e^{-n_\text{total}} \prod_{j=1}^{N_\text{bins}} \frac{1}{C_j!}\big[n_\nu P_{\nu,j} + n_c
P_{c,j} + n_b P_{b,j} \big]^{C_j},
\end{equation}

\noindent where $C_j$ is the observed number of events in bin $j$, $n_\nu$, $n_c$, and $n_b$ are
the fitted number of $\cevns$, capture, and background events, respectively, $n_\text{total}$ is
the sum of these, and $P_{\nu,j}$, $P_{c,j}$, and $P_{b,j}$ are the probability of an event of
each type lying in bin $j$ (all normalized such that they sum to unity over all bins).

We employ a likelihood ratio test on $\mathcal{L}/\mathcal{L}'$, where $\mathcal{L}'$ is a similar
likelihood function but without the \cevns\ contribution--with the parameter $n_\nu$ in
Eq.~(\ref{eqn:likelihood_alt}) set to zero.

The two likelihoods are compared using Wilks' theorem, which states that
$2\ln(\mathcal{L}/\mathcal{L}')$ is distributed as a chi-squared random variable with, in this
case, one degree of freedom~\cite{Wilks:1938dza}. This leads immediately to a probability for
comparing the nested models, i.e., the probability that the combined model of
Eq.~(\ref{eqn:likelihood_alt}) \textit{is} a better model than the simple model with
$n_\nu$$\equiv$0.  We quantify the \textit{confidence level} as the Z-value ($\sigma$ value)
corresponding to the symmetric (two-sided) normal error integral with the same probability, i.e.,
$\sqrt{2\ln(\mathcal{L}/\mathcal{L}')}$. 

\begin{figure*}[htb]
    \centering
    \includegraphics[width=2\columnwidth]{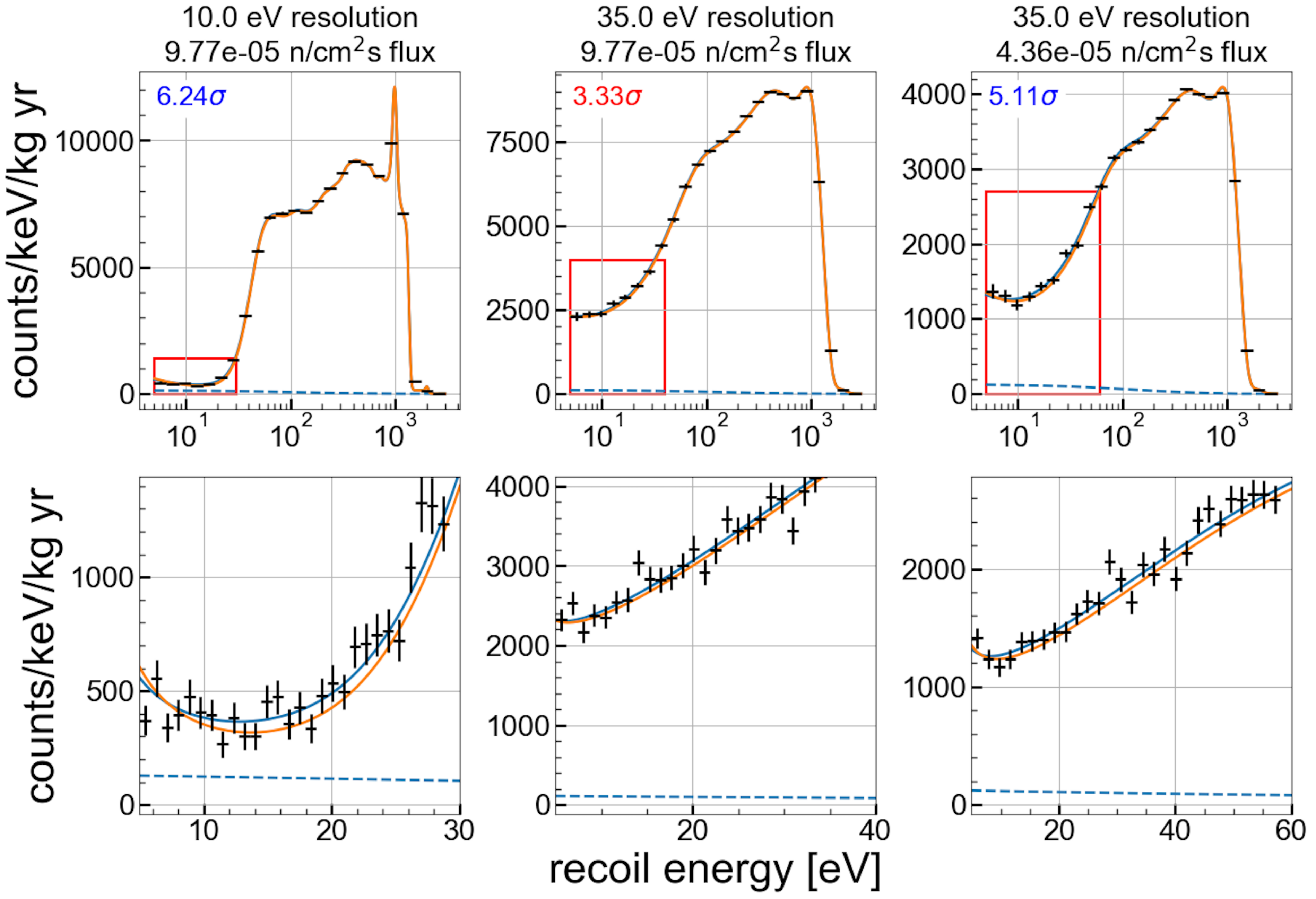}
    \caption{(Color online) Likelihood fitting examples for three different combinations of
resolution and neutron flux for $\sigma_0=$1\,eV in silicon. The left two plots show a flux of
$9.77\times10^{-5}$~\fluxunits, corresponding to $\sim1.1$\,M capture events, and the right is
lower flux, $4.36\times10^{-5}$~\fluxunits, corresponding to $\sim500,000$ capture events. The
right two plots show data with an effective resolution of 35\,eV, and the left plot shows an
effective resolution of 10\,eV.  The black points show the histogram of our toy model data for
given \cevns, thermal flux, and background contributions.  Solid blue and orange curves show the
fitted PDFs for with- and without-\cevns\ fits. The dashed blue curve shows the \cevns\
contribution to the with-\cevns\ fit.  
}
    \label{fig:fit_examples}
\end{figure*}

Figure~\ref{fig:fit_examples} shows example fits for different combinations of effective
resolutions and neutron fluxes in silicon. Resolutions of 10\,eV and 35\,eV are used in conjunction
with neutron fluxes of $4.36\times10^{-5}$~n/cm$^2$s and $9.77\times10^{-5}$~n/cm$^2$s with $\sigma_0=1$\,eV.
The black points show the histogram of our toy model data for given \cevns, thermal flux, and
background contributions. The toy model data is smeared using our hypothetical resolution
functions and plotted for recoil energies between 0\,eV and 1800\,eV. Solid blue and orange curves
show the fitted PDFs for with- and without-\cevns\ fits. The dashed blue curve shows the \cevns\
contribution to the with-\cevns\ fit.  The confidence level of each fit is given in the upper left 
corner of each plot, red if less than $5\sigma$ and blue if greater. 

\begin{figure*}[htb]
    \centering
    \includegraphics[width=2\columnwidth]{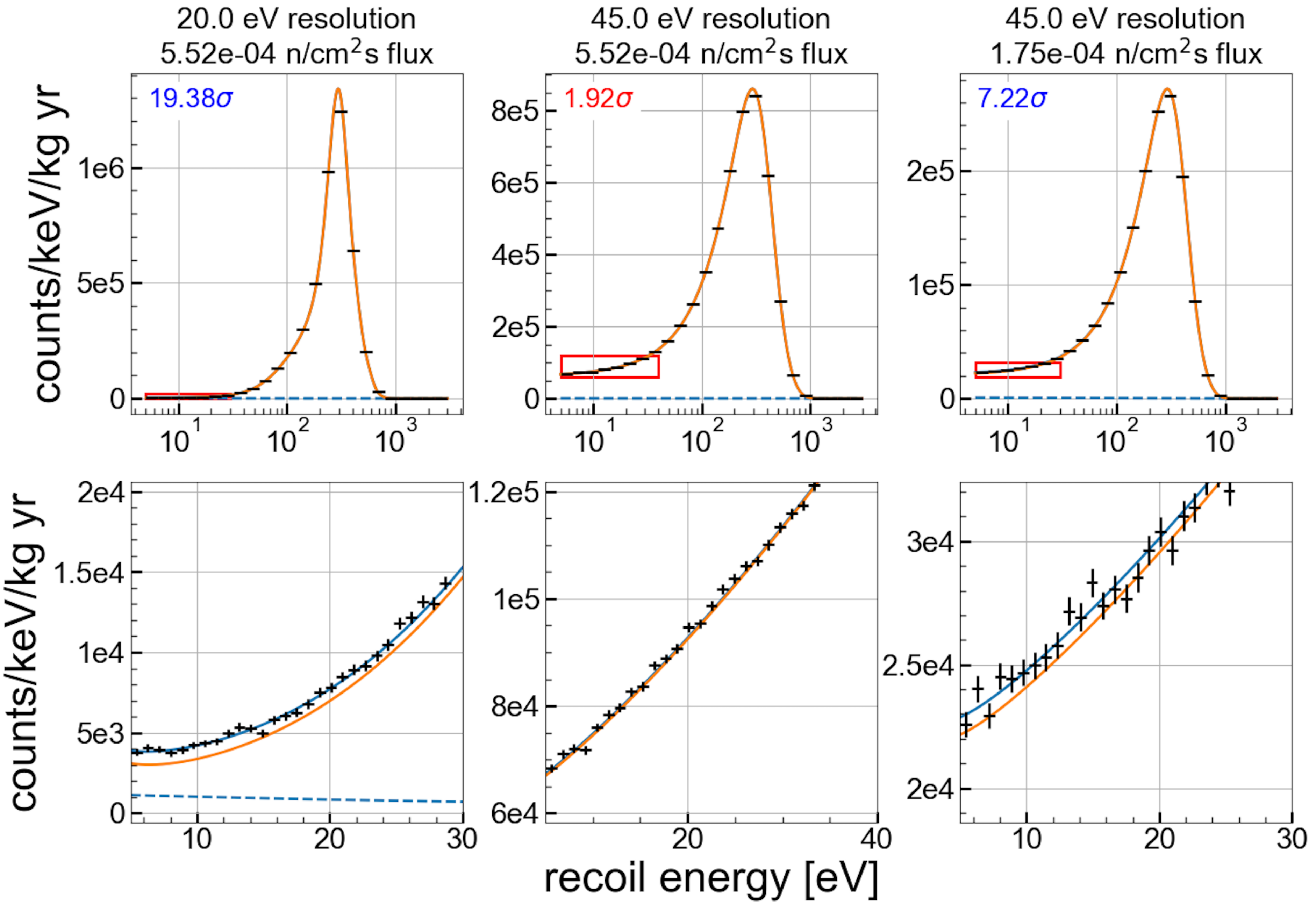}
    \caption{(Color online) Likelihood fitting examples for three different combinations of
resolution and neutron flux for $\sigma_0=$1\,eV in germanium. The left two plots show a flux of
$5.52\times10^{-4}$~\fluxunits, corresponding to $\sim 31.6$\,M capture events, and the right is
lower flux, $1.75\times10^{-4}$~\fluxunits, corresponding to $\sim 10$\,M capture events. The
right two plots show data with an effective resolution of 45\,eV, and the left plot shows an
effective resolution of 20\,eV.  The black points show the histogram of our toy model data for
given \cevns, thermal flux, and background contributions.  Solid blue and orange curves show the
fitted PDFs for with- and without-\cevns\ fits. The dashed blue curve shows the \cevns\
contribution to the with-\cevns\ fit.  
}
    \label{fig:fit_examples_Ge}
\end{figure*}

In Figure~\ref{fig:fit_examples} the leftmost and rightmost plots show the two methods by which the
\cevns\ signal can be detected, either by improving resolution so the \cevns\ signal can be
discerned on the low-energy downslope (leftmost plot), or by reducing the neutron flux so that the
peak caused by \cevns\ events is clearly visible over the signal from capture events (rightmost
plot). The center plot shows a scenario where the thermal neutron flux is too high to produce a
5$\sigma$ \cevns\ detection with the assumed livetime. Figure~\ref{fig:fit_examples_Ge} shows the
analogous fit plots for germanium. In the germanium case we use resolutions of 20\,eV and 45\,eV
with thermal neutron fluxes of $5.52\times10^{-4}$~n/cm$^2$s and $1.75\times10^{-4}$~n/cm$^2$s.
The thermal neutron fluxes used for germanium are generally higher than for silicon because
germanium is more resilient to the flux --- although there are more capture events the capture spectrum overlaps less strongly than in silicon.

\clearpage
\section{\label{sec:conclusions}Results and Discussion}

\begin{figure*}[htbp]
    \centering
    \includegraphics[width=2\columnwidth]{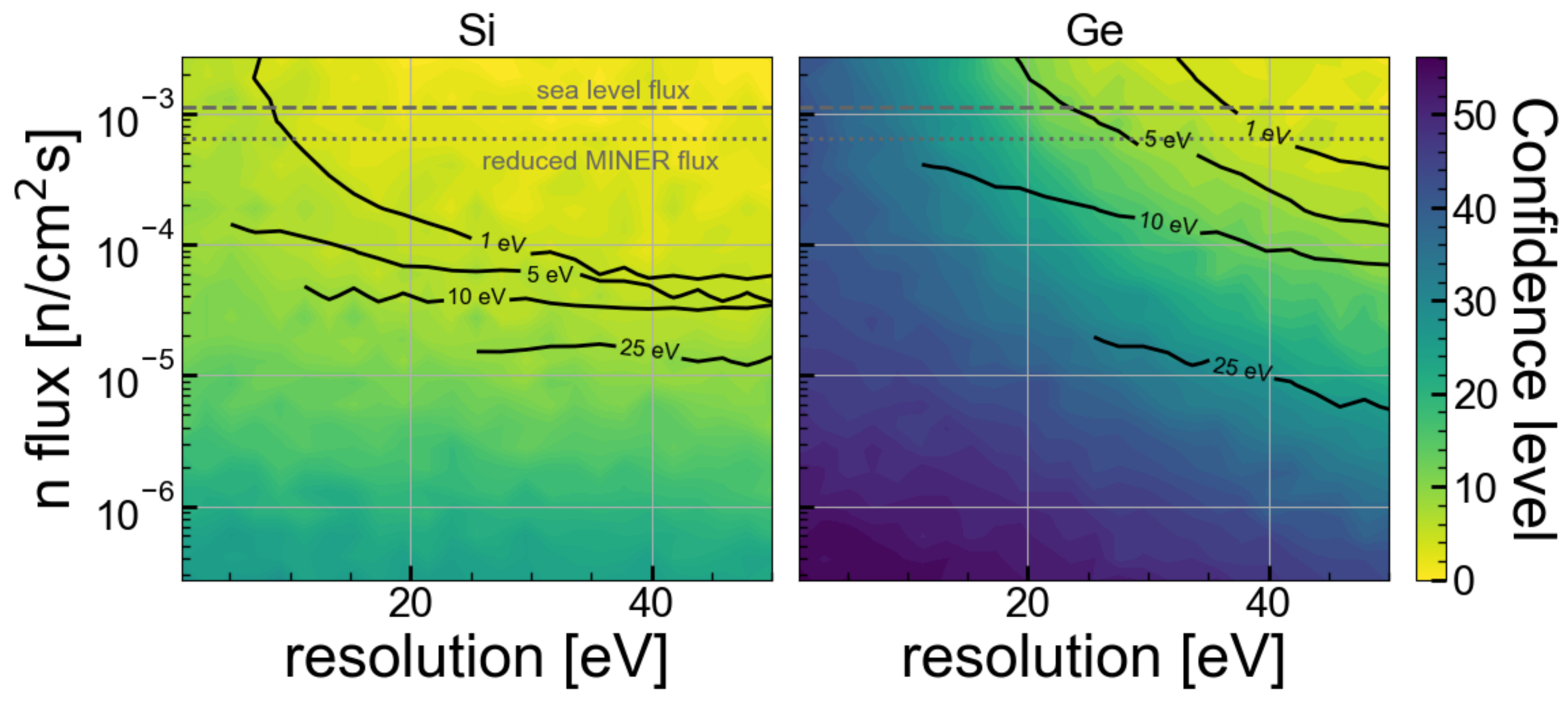}
    \caption{(Color online) Confidence level for varying ambient thermal neutron flux (vertical axis) and effective
resolution (horizontal axis) for a baseline resolution of 1\,eV in Si (left) and Ge (right).
Horizontal lines mark fluxes corresponding the accepted ambient sea level flux (dashed) and one part
in $10^{9}$ of the measured flux at the MINER facility~\cite{AGNOLET201753} (dotted). Black contour
lines representing $5\sigma$ confidence levels for different values of the baseline resolution
$\sigma_0$ are also shown. The value of $\sigma_0$ corresponding to each contour is given by the
label on each contour. } \label{fig:confidences}
\end{figure*}

Figure~\ref{fig:confidences} shows the results for 100 kg\,yr exposure in Si (left) and Ge (right)
for $\sigma_0 = 1$\,eV. To produce this plot, we generated a grid of 25 resolution values and 25
flux values and completed fits on that grid for each material. The contour plot is a smoothing of
those fit results. The vertical axis shows ambient thermal neutron flux and the horizontal axis
shows the effective resolution. Colors correspond to confidence levels. The two horizontal broken 
lines delineate a few reference flux levels --- the gray dashed line shows the generally accepted ambient
neutron flux at sea level, 4~cm$^{-2}$\, hr$^{-1}$~\cite{1263842}. The gray dotted line shows 1 part in
$10^{9}$ of the neutron flux measured at the MINER facility,
5.8$\times10^{7}$~cm$^{-2}$\,s$^{-1}$~\cite{AGNOLET201753}, extrapolated to ten meters from the
reactor core. As reference, the two lines correspond to approximately 13,000,000 (63,000,000) and
7,500,000 (37,000,000) neutron capture events in Si (Ge), respectively.

The black lines represent $5\sigma$ contours, approximately 99.99994\% confidence in the presence
of the \cevns\ signal, for three different values of $\sigma_0$ (the 1\,eV contour is the contour
for the pictured colormap data).  Each contour was yielded after smoothing the data by six
iterations through a Jacobi relaxation scheme. Above each line, it is unlikely that the
\cevns\ signal can be extracted from the data given that baseline resolution.

Up to statistical fluctuations, the confidence increases monotonically moving toward lower fluxes
and better resolutions, as would be expected. This is the case for all baseline resolutions
probed. 

Many interesting features can be seen in this data. For $\sigma_0=1$\,eV, below an effective
resolution of $\sim20$\,eV, improving resolution has a dramatic effect on tolerance to ambient
neutrons. An improvement in effective resolution from 10\,eV to 20\,eV is enough to allow measurement
of \cevns\ in an ambient flux three times larger. However, at larger baseline resolutions, the
improvement with effective resolution is much smaller, and at $\sigma_0=10$~eV, changing the
effective resolution does almost nothing. This is an instructive, and promising, result,
emphasizing that the feasibility of \cevns\ measurements gets ever greater as detector
technologies improve, but that decreasing the thermal neutron flux of current experiments may be a
vital step in achieving these measurements. 

Further, even with drastic detector improvements, thermal neutrons could still represent a
substantial background in the vicinity of nuclear reactors. It can be seen in Figure
\ref{fig:confidences} that the measured MINER reactor-adjacent flux needs to be brought down to
one part in a billion to begin seeing a \cevns\ signal at a baseline resolution of only 1\,eV, and
must be brought down even further at $\sigma_0=10$\,eV. Note that the flux measurement by
MINER~\cite{AGNOLET201753} was early in the campaign. Many materials like boron, cadmium,
gadolinium, polyethylene, and more can be used to bring down thermal fluxes by large factors. The
thermal neutron background situation is somewhat better in germanium detectors owing to the larger
\cevns\ cross section and lower capture rates at low energy. 

We have done our analysis in terms of nuclear recoil energies but many detectors will measure the
ionization caused by nuclear recoils instead. This will tend to shift all the nuclear recoil
events toward lower energies, as the ionization yield is below unity and typically lower at lower
energies~\cite{osti_4701226}. However, we expect our results to be unaffected as the shift is
largely the same for both capture and neutrino induced recoils. 
The resolution and detector threshold would also have to be adapted to \hlh{the specific case.}

In germanium detectors the thermal neutron flux can be measured \emph{in situ} and compared with
our results. The flux measurement is based on the rates of one of the X-ray emissions from
electron capture of $^{71}$Ge. After several half-lives ($T_{1/2}$=11.4\,days) the ambient neutron
flux will be equal to the rate of the $K$-shell electron capture (EC) line divided by a quantity
which is the product of the branching of that line, the macroscopic cross section of $^{70}$Ge
capture, and the volume of the detector. 


Thus far, this study has not discussed the possibility of vetoing events based on outgoing capture
gamma rays. In general, capture events are highly vetoable by looking for signs of the gammas
emitted during each capture. As a simple estimate of plausible effective flux reduction,
Table~\ref{tab:veto_reductions} lists the expected effective flux $f_\text{effective}$ over the
actual flux $f_\text{actual}$ as a function of $\bar{d}$, the angle-weighted average distance an
emitted gamma would travel before leaving a detector made of either silicon or germanium. Values
are based on the probability of gammas from capture event leaving the detector without
interacting, based on the same cascades (potentially involving multiple exiting gamma rays) and
cascade rates used to generate the capture event spectra, and using gamma/nucleus cross sections
yielded via linearly interpolating data from NIST's XCOM database~\cite{10.18434/T48G6X}. These
estimates show that physically larger detectors can make great gains in reducing the importance of
ambient thermal neutrons, particularly in the case of germanium.

\begin{table}[htbp]
    \centering
    $f_\text{effective}/f_\text{actual}$
    
    \begin{tabular*}{0.8\columnwidth}{c @{\extracolsep{\fill}} c c}
        \hline \\[-2ex]
        $\bar{d}$ (cm) & Si  & Ge\\
        \hline
        1 & 0.8382 & 0.442 \\
        5 & 0.4267 & 0.07027 \\
        10 & 0.1937 & 0.01251 \\
        20 & 0.04538 & 0.001166 \\
        50 & 0.001747 & 7.313e-06 \\
        100 & 7.119e-05 & 1.917e-09 \\
        \hline
    \end{tabular*}
    \caption{Expected effective thermal neutron flux $f_\text{effective}$ over actual flux
$f_\text{actual}$ in silicon and germanium detectors as a function of $\bar{d}$, the average
distance a gamma must travel before leaving the detector. Reduction is based on the probability of
gammas leaving the detector without interacting.} \label{tab:veto_reductions}
\end{table}

Some assumptions about the ``other'' backgrounds should be noted. The low-energy-skewed spectrum
(${\propto E^{-0.9}}$) can be viewed as an instantiation of the anomalous excess of events \hlh{in low-threshold cryogenic detectors} at low
energy recently measured by several collaborations~\cite{10.21468/SciPostPhysProc.9.001,
AGNOLET201753, Aguilar-Arevalo_2016, PhysRevD.100.102002, PhysRevLett.125.241803,
PhysRevD.99.082003, PhysRevLett.125.141301, ARNAUD201854, Rothe2020, Billard_2017,
PhysRevLett.125.171802, PhysRevD.102.091101, PhysRevLett.127.061801, PhysRevLett.121.051301}. The
origin of these events are unclear, \hlh{and a single common explanation for the excesses is unlikely, making it difficult to assess the likely form for such an excess in our simulated experiment.}
However, since we consider the nuclear recoil energy scale
these excess events are either similarly peaked to our assumed additional background model, or
they are more spread out due to the less-than-unity quenching factor of nuclear recoils.

The last point that bears discussion is that the reactor model used here may be a best-case
scenario. A parallel track of calculations in silicon using antineutrino spectrum data from the
Daya Bay nuclear reactor complex in Southern China~\cite{An_2017} yielded consistently lower
confidence values. This fact highlights possible systematic uncertainties in our analysis due to
uncertainties in the reactor anti-neutrino spectrum. The precise shape and overall normalization
of this spectrum is actively being
improved~\cite{doi:10.1146/annurev-nucl-102115-044826,PhysRevLett.123.022502}, but the current
status does not preclude changes to our results based on the form of the spectrum. This is
expected to be a more significant effect than changes based on uranium enrichment and relative
fission contributions from $^{235}$U, $^{238}$U, $^{239}$Pu, and $^{241}$Pu. This is certainly a
less significant effect than the ability to veto capture events, but still should be noted.

Overall, our analysis shows that keeping the effective thermal neutron flux (flux after veto)
below $\sim$~10$^{-4}$\,\fluxunits\ for a 1\,MW reactor at 10 meters is necessary for silicon
detectors--even ones with exceptional resolutions.  Germanium is somewhat more robust for
detectors with exceptional resolution, requiring the flux to be below
$\sim$~7$\times$10$^{-4}$\,\fluxunits\ under the same conditions. For germanium detectors with
25\,eV baseline resolution (still excellent by today's standards) the thermal neutron flux needs
to fall below $\sim$~10$^{-5}$\,\fluxunits.  These rules of thumb can be roughly adapted to other
reactors by multiplying this flux by the reactor power in MW (uranium enrichment level probably
makes a sub-dominant difference) and accounting for the distance by an inverse-square law for the
neutrinos. 

All of the Python code used for this analysis can be obtained via the Open Storage Framework (OSF)
data entry~\cite{Biffl_Harris_Villano_2023}. The PDFs for the thermal neutron-induced nuclear
recoils were obtained from our public code~\cite{Villano2022}.

\begin{acknowledgements}

We gratefully acknowledge support from the U.S. Department of Energy (DOE)
Office of High Energy Physics and from the National Science Foundation (NSF).
This work was supported in part by DOE Grant DE-SC0021364 and NSF Grant
No.~2111090.  
\end{acknowledgements}

\clearpage
\bibliography{prdrefs_short.bib}
\bibliographystyle{apsrev4-2}

\end{document}